\documentstyle[preprint,aps,epsfig]{revtex}

\newcommand{\be}{\begin{eqnarray}}
\newcommand{\ee}{\end{eqnarray}}
\renewcommand{\L}{\bar{\Lambda}}

\begin{document}

\draft
\preprint{
YUMS 97--14
\hspace{-36.5mm}
\raisebox{5.0ex}{
KAIST-TH-97/08
}
\hspace{-32.0mm}
\raisebox{2.5ex}{SNUTP 97--069} 
}

\title{On the Motion of $b$-quark inside $B$ meson:\\
the Parton Model~ {\it vs.} the ACCMM Model\\}
\author{
C. S. ~Kim$^{a}$\thanks{kim@cskim.yonsei.ac.kr,~~ cskim@kekvax.kek.jp},
Yeong Gyun ~Kim$^b$
\thanks{ ygkim@chep6.kaist.ac.kr} and
Kang Young ~Lee$^a$\thanks{kylee@chep5.kaist.ac.kr}
\thanks{ Department of Physics, KAIST, Taejon 305--701, Korea }
}
\vspace{0.5cm}
\address{
{\footnotesize a}
Department of Physics, Yonsei University, Seoul 120--749, Korea\\
{\footnotesize b}
Department of Physics, Korea Advanced Institute of 
Science and Technology, Taejon 305--701, Korea
}

\date{\today}
\maketitle
\begin{abstract}

We study the distribution function of the $b$--quark
for inclusive semileptonic $B$ decays in the framework
of the parton model.
We compare the functional behavior of the fragmentation
inspired function and the Gaussian ACCMM function,
and discuss their validity.
Using the HQET parameters,
we can critically test these models.

\end{abstract}

\pacs{ }

\narrowtext

\section{Introduction}

Studies of inclusive semileptonic decays of $B$ mesons have drawn 
our interest as a testing ground for the motion of $b$--quarks
inside hadrons.
We have the differential decay width for the process
$B \to X_q l \nu$
\be 
d\Gamma \sim W^{\mu \nu} L_{\mu \nu}
                               \frac{d^3 p_l}{(2\pi)^3 2 E_l}
                               \frac{d^3 p_\nu}{(2\pi)^3 2 E_\nu},
\ee
where $L_{\mu \nu}$ is the leptonic tensor and 
the hadronic tensor $W^{\mu \nu}$ is defined as
\be
W^{\mu \nu} = \frac{1}{2} \sum_{spins} 
     \int \frac{d^3 p_{_X}}{(2\pi)^3 2E_X} (2\pi)^4 \delta^4 (p_{_B} - q -p_{_X})
     \langle B | {J^\mu}^{\dagger} | X \rangle
     \langle X | J^\nu | B \rangle .
\ee
Since the $B$ meson is heavy, the momentum transfer of the decay is
larger than the typical energy scale of hadronic binding, 
$\sim \Lambda_{QCD}$, in most regions of phase space.
This enables us to formulate the inclusive semileptonic decay process
in an analogous manner to the well--known method of the deep inelastic
scattering (DIS).
Light--cone dynamics dominates the decay process and the hadronic 
form factors show the remarkable feature of scaling 
in the absence of the relevant corrections.
So the parton picture should work well, where the modeling
of the decay process is possible by 
summing the partonic decay processes weighted by
the distribution of the $b$--quark momentum inside the $B$ meson.
Bareiss and Paschos \cite{bareiss} noticed this point and they formulated
the inclusive decays along the lines of analyzing DIS in the parton picture.
The general expression for the hadronic tensor in terms of 
invariants is given by
\be
W^{\mu \nu} 
(q^2, q \cdot p_{_B}) 
&=& -g^{\mu \nu} W_1 (q^2, q \cdot p_{_B}) 
+ p_{_B}^{\mu} p_{_B}^{\nu} \frac{W_2 (q^2, q \cdot p_{_B}) }{m_{_B}^2}
\nonumber \\
&&- i \epsilon^{\mu \nu \alpha \beta} {p_{_B}}_{\alpha} q_{\beta}
  \frac{W_3 (q^2, q \cdot p_{_B})}{2 m_{_B}^2}
+ q^{\mu} q^{\nu} \frac{W_4 (q^2, q \cdot p_{_B})}{m_{_B}^2}
\nonumber \\
&& + (p_{_B}^{\mu} q^{\nu} + p_{_B}^{\nu} q^{\mu}) 
    \frac{W_5 (q^2, q \cdot p_{_B})}{2 m_{_B}^2} 
- (p_{_B}^{\mu} q^{\nu} - p_{_B}^{\nu} q^{\mu}) 
    \frac{W_6 (q^2, q \cdot p_{_B})}{2 m_{_B}^2} .
\ee
Following the parton picture of Feynman \cite{feynman},
we let the probability of finding a $b$-quark in a $B$ meson 
carrying a fraction $x$ of the meson momentum 
in the infinite momentum frame
be the distribution function $f(x)$.
The decay width is then obtained  by incoherently summing
the partonic decay widths weighted by $f(x)$ in that frame.
Thus we immediately find that
\be
&W_1 (q^2, q \cdot p_{_B}) = \frac{1}{2} \left( f(x_+) + f(x_-) \right)
\nonumber \\
&W_2 (q^2, q \cdot p_{_B}) = \frac{2}{x_+ - x_-} 
                         \left( x_+ f(x_+) - x_- f(x_-) \right)
\nonumber \\
&W_3 (q^2, q \cdot p_{_B}) = - \frac{1}{x_+ - x_-} 
                           \left( f(x_+) - f(x_-) \right)
\nonumber \\
&W_{4,5,6} (q^2, q \cdot p_{_B}) = 0
\ee
after neglecting the lepton masses
where 
\be
x_{\pm} = \frac{q_0 \pm \sqrt{{\bf q}^2+m_q^2}}{m_{_B}}
\ee
where $m_q$ is the final state quark mass in $b \to qe\nu$.
This approach has been reformulated on the light--cone 
in Ref. \cite{jp,jin0,rad} and holds in any frame.
The QCD radiative correction of the parton model was studied 
in Ref. \cite{rad}.
The idea of the parton model was also applied to 
the exclusive semileptonic decays of $B$ meson 
in Ref. \cite{cskim1}.

Recently the heavy quark effective theory (HQET) has been applied to
inclusive semileptonic decays of $B$ mesons 
and brought great progress \cite{chay,manohar,bigi,bigi0,neubert}.
With the expansion for the decay rate in inverse power of 
$b$--quark mass, a systematic study is possible
and one obtains more information about the hadronic
matrix elements than before.
Perhaps one of the most important results of the HQET analysis
for the inclusive decays is the absence of the contribution of the $1/m_b$
correction terms. 
With the help of the $1/m_b$ expansion, we can relate
the moments of the charged lepton energy spectrum
to the matrix elements for the operator expansion
of the current product. 

The $1/m_b$ structure of the inclusive parton model 
has recently been analyzed in Ref. \cite{kylee}.
When the parton model approach motivated by DIS is applied 
to the analysis of $B$ decays,
the modeling boils down to introducing the distribution function
of Eq. (4), which describe the $b$--quark's motion inside the $B$ meson.
In other words, in the study of the inclusive semileptonic $B$ decays,
the problem of which model we choose corresponds to
the problem of choice of the distribution function.
In general two models are known which predict
the lepton energy spectrum of inclusive decays of $B$ meson.
Altarelli et al. (ACCMM) \cite{accmm} treat the $b$--quark and the
spectator quark as quasi--free particles with a Gaussian
spectrum of Fermi momentum in the $B$ rest frame.
Bareiss and Paschos \cite{bareiss,jin} notice that the $b$--quark
distribution inside a $B$ meson in the infinite momentum frame (IMF)
is related to the fragmentation
of a $b$--quark into a $B$ meson by crossing symmetry of the spectator, 
and so their functional behaviors are similar to each other.
Thus they chose Peterson's fragmentation function \cite{peterson}
as the distribution function of $b$--quark, which is 
the default choice in Lund Monte Carlo programs.
Both models get results at the hadronic level after convolution
of the quark level spectrum.
The difference between the two models mainly comes from 
the weight function which is corresponding to 
the distribution function $f(x)$ of $b$--quark momentum
in the parton model approach.

In this paper, we study both models 
from the viewpoint of the parton model picture
to compare their distribution functions with each other. 
For explicit comparison of the distribution functions 
of both models, we derive the distribution functions
described by the same definition in the same reference frame.
Moreover, with the help of the $1/m_b$ study suggested in Ref. \cite{kylee}
we attempt to test the models by extracting the HQET parameters,
the effective mass of light degrees of freedom $\L$ 
and the average kinetic energy of 
$b$--quark $\mu_\pi^2$.
We show that known bounds on $\L$ and $\mu_\pi^2$ constrain
the model parameters and may even rule out the model itself.
In section 2, we review the parton model approach first. 
In this approach Peterson's function and its improved form
by Lee and Kim \cite{kylee} is taken as the distribution function.
In the ACCMM model, the distribution of the $b$--quark is
determined by the Gaussian distribution for the Fermi motion
of the spectator quark in the $B$ meson rest frame.
In section 3, we derive the distribution function of the ACCMM model 
in the infinite momentum frame from the Gaussian function,
which allows a consistent comparison of the two models.
The behavior of the distribution functions
is investigated and the heavy quark effective theory (HQET)
parameters are obtained in terms of the parameters of each model. 
Our discussions and conclusions are summarized in section 4.

\section{The Parton Model}

We first review the parton model approach
applied to analyze inclusive semileptonic $B$ meson decays.
As described in the previous section,
the distribution function $f(x)$ is defined as 
the probability of finding a $b$-quark in a $B$ meson 
carrying a fraction $x$ of the meson momentum 
in the infinite momentum frame.
Thus the decay probability is determined by the momentum
of the $b$--quark and has the value of $f(x)dx$.
And we write the Lorentz invariant decay width as follows
\be
E_B~d\Gamma(B \to X_q e \nu) = \int~dx~f(x)~
                 E_b~d\Gamma(b \to qe\nu) ,
\ee
with the relation $p_{_b} = x p_{_B}$.
Since $B$ mesons are heavy, light--cone dynamics dictates
the inclusive semileptonic decay analogous to that of DIS.
The universal distribution function of QCD is interpreted
as the structure function of the $b$--quark in the parton model,
which determines the distribution of the light--cone projection
of the $b$--quark momentum inside the $B$ meson.
We directly derive the triple differential decay rate as
\be
\frac{d\Gamma}{dE_l dq^2 dq_0} = \frac{G_F^2 |V_{qb}|^2}{4\pi^3}
            \frac{q_0-E_l}{\sqrt{{\bf q}^2+m_q^2}}
            \left[ x_+ f(x_+) (2E_l-m_{_B}x_-)
                    + (x_+ \leftrightarrow x_-) \right]
\ee
where $x_\pm$ is defined in the previous section.
Jin and Paschos \cite{jp} argued that the contribution
of $f(x_-)$ is expected to be relatively small in the kinematic
region we are interested in, and can be neglected.
This is a consequence of the fact that
the light--cone distribution function is sharply peaked
around $x \sim m_b/m_{_B}$.

For a more quantitative analysis, 
we have to introduce a method analogous to the HQET.
In this parton model approach, information on the nonperturbative
dynamics inside the $B$ meson is encoded in the distribution function.
Therefore, it is crucial to determine precisely the form of the distribution 
function in this approach. 
Unfortunately we don't presently know how to calculate the true form of it. 
However, we have some knowledge about the line shape function
of QCD from the help of HQET and the operate product expansion (OPE).
Since the mass of the $b$--quark does not appear 
in this parton model framework,
we have to define it within this framework in order to express the model
in terms of the inverse power expansion of the $b$--quark mass.
With properly defined $m_b$, the parton model is well represented by 
an $1/m_b$ expansion \cite{kylee}.
We define
\be
m_b \equiv \langle x \rangle m_{_B} ,
\ee
where $\langle x \rangle = \int dx~x f(x)$. 
With this definition as the starting point,
we can relate the parton model to HQET and compute 
the effective mass of the light degrees of freedom in a $B$ meson, $\L$.

In the HQET framework, the triple differential decay rate for 
$B \to X_q l \nu$ shows scaling behavior
\cite{manohar,bigi,bigi0}:
\be
\frac{d\Gamma}{dE_l dq^2 dq_0} = \frac{G_F^2 |V_{qb}|^2}{192\pi^3}
            \frac{2}{\L} F(x_{_B})
            \frac{24(q_0-E_l)(2m_bE_l-q^2)}{m_b-q_0} ,
\ee
where
\be
x_{_B} = -\frac{m_b^2+q^2-2 m_b q_0}{2 \L (m_b-q_0)} .
\ee
Apart from kinematic factors, the triple differential decay
rate is completely determined by one scaling variable $x_{_B}$.
The scaling behavior of QCD will be violated by perturbative
and nonperturbative (higher twist) corrections. 
If we take the final state quark to be $u$-quark and so $m_q=0$
hereafter, the scaling form is derived from Eq. (6)
\be
\frac{d\Gamma}{dE_l dq^2 dq_0}
\approx \frac{G_F^2 |V_{ub}|^2}{192\pi^3}
            \frac{2}{m_{_B}} f(x_+)
            \frac{24(q_0-E_l)(2m_bE_l-q^2)}{m_b-q_0}
\ee
where $m_{_b}$ is as defined in Eq. (7).
{}From Eqs. (8) and (10), 
the line shape function $F(x_{_B})$ of QCD is related to 
the distribution function $f(x_+)$ 
of the parton model after appropriate variable changes
\be
\frac{1}{m_{_B}} f(x_+) =
\frac{1}{\L} F(x_{_B}) .
\ee

Next we intend to study their functional behavior.
Following the notation of Bigi et al. \cite{bigi,bigi0},
a general form for the singular expansion of 
the distribution function is written as
\be
F(x_{_B}) = \sum_{n=0}^{\infty} \frac{(-1)^n}{n!} a_n \delta^{(n)} (x_{_B}) ,
\ee
where moments of the equation:
\be
a_n = \int~dx_{_B}~x_{_B}^n F(x_{_B}) ,~~~~~~n=0,1,\cdot \cdot \cdot ~.
\ee
Using the HQET and the OPE, we parametrize the moments with 
the forward scattering matrix elements of $B$ meson.
Especially we know that $a_0=1$ from $b$--number conservation,
and that $a_1=0$ up to ${\cal O}(1/m_b^2)$,
which indicates the lack of ${\cal O}(1/m_b)$ corrections.
The value of the first moment $a_1=0$ is related to the value of 
the $b$-quark mass $m_b$ or the parameter $\bar{\Lambda}$. 

The second moment of the spectrum is related to
the average kinetic energy of $b$--quark inside the $B$ meson
and usually represented by the parameter $\mu_\pi^2$, defined by
\be
a_2 = \frac{1}{3\L^2} \frac{1}{2m_{_B}}
           \langle B| \bar{b} {\bf \pi}^2 b |B \rangle
    \equiv \frac{\mu_{\pi}^2}{3\L^2} ,
\ee
where ${\bf \pi}_\alpha = i D_\alpha - m_b v_\alpha$ 
with $D_\alpha$ the covariant derivative and  
$v_\alpha$ the four--velocity of the $B$ meson.
After the variable changes of Eq. (11), we can calculate the moments
for the distribution function in the framework of the parton model.
Under the definition of $m_b$ in Eq. (7), we have $a_1 = 0$.
The second moment is derived as
\be
a_2 = \int^1_{-\infty}~dx_{_B}~x_{_B}^2 F(x_{_B})
    = \left( \frac{m_{_B}}{\L} \right)^2
        \int^1_0~dx~ ( x-\frac{m_b}{m_{_B}} )^2 f(x)
    = \left( \frac{m_{_B}}{\L} \right)^2
      (\langle x^2 \rangle -\langle x \rangle^2) .
\ee
We first used Peterson's fragmentation function as the distribution
function of $b$--quark 
following the original work of Bareiss and Paschos \cite{bareiss}, 
and obtained the value 
for $a_2 \sim 0.76$, with the model parameter $\epsilon = 0.004$
and the mass of $B$ meson $m_{_B}=5.3$ GeV,
which agrees well with the QCD sum rule result of Ref. \cite{ball}.
As pointed out in the Ref. \cite{kylee}, however, 
this function gives rather larger $\bar{\Lambda}$ and $\mu_{\pi}^2$,
and we need to find a better functional form or some modification
for Peterson's function.
Jin and Paschos \cite{jp} also pointed out that the distribution function
should be more sharply peaked than Peterson's function. 
By improving Peterson's arguments, Lee and Kim \cite{kylee} suggested a 
modified form of Peterson's function 
\be
f^{\mbox{{\tiny new}}}_Q(x) = \frac{N_Q}
             {x(1+\alpha_{_Q}-\frac{1}{x}-\frac{\epsilon_{_Q}}{1-x})^2} .
\ee
Using this new form, we calculate
$\bar{\Lambda}$ and $\mu_{\pi}^2$  by varying the input parameters
$\alpha_{_Q}$ and $\epsilon_{_Q}$.

Fig. 1 shows the behavior of the distribution function
with varying $\alpha_{_Q}$ and $\epsilon_{_Q}$.
The larger the value of $\alpha_{_Q}$, the sharper the functional behavior.
And the smaller the value of $\epsilon_{_Q}$, 
the sharper the function and the closer the location of peak to 1.
We also show the parameter space of 
$(\alpha_{_Q} , \epsilon_{_Q})$
for allowed $\bar{\Lambda}$ and $\mu_{\pi}^2$ in Fig. 2.
For the values of $\mu_{\pi}^2$, there is
some theoretical controversy which we will discuss later in section 4.
We take the range of $\L$ and $\mu_\pi^2$ obtained by 
the HQET and QCD sum rules in Refs. \cite{ball,neubert1}
\be
\L = 0.4 \sim 0.6,
\nonumber 
\ee
\be
\mu_{\pi}^2 = 0.6 \pm 0.1~~\cite{ball},~~~~{\rm and}~~~~
\mu_{\pi}^2 = 0.10 \pm 0.05~~\cite{neubert1}.
\ee
Ball and Braun \cite{ball} calculated $\mu_\pi^2$ using the QCD sum rule
approach and obtained $\mu_\pi^2= 0.60 \pm 0.10$ ${\rm GeV}^2$
for $B$-meson, while Neubert \cite{neubert1} obtained 
$-\lambda_1= 0.10 \pm 0.05$ ${\rm GeV}^2$.
We find that there exists a region in the 
$(\alpha_{_Q} , \epsilon_{_Q})$ plane that gives values
of $\L$ and $\mu_\pi^2$ in agreement with those of Ball and Braun \cite{ball}.
If Neubert's prediction \cite{neubert1} for the value of $\mu_\pi^2$ is correct,
there is no common region in which the model parameters
give the values of $\L$ and $\mu_\pi^2$ in Eq. (17),
and we may conclude that this functional form
is not appropriate for inclusive semileptonic $B$ decays. 

\section{The ACCMM Model}

The ACCMM model \cite{accmm} considers the $B$ meson as consisting of
a $b$-quark and a spectator quark. 
The spectator quark is treated as a quasi--free particle having 
definite mass $m_{sp}$, and momentum ${\bf p}$, 
while the $b$--quark is treated as a virtual particle of
invariant mass $m_b$
\be
m_b^2(p)=m^2_B+m^2_{sp}-2m_{_B}\sqrt{p^2+m^2_{sp}} ,
\ee
where $m_{_B}$ is the $B$ meson mass and $p \equiv |\bf{p}|$
in the $B$ meson rest frame.
The $b$--quark energy is given by
\be
E_b(p)&=&
\sqrt{p^2+m_b^2(p)} \\ \nonumber
&=&m_{_B}-\sqrt{p^2+m^2_{sp}} ~.
\ee
The distribution of the spectator momentum is usually
assumed to be a Gaussian with an adjustable parameter $p_{_F}$
\be
{\Phi}(p)=\frac{4}{{\sqrt{\pi}} p_{_F}^3}~ 
            \exp \left(-\frac{p^2}{p_{_F}^2}\right)
\ee
with the normalization
\be
\int_{0}^{\infty} dp~ p^2 \Phi(p) = 1 .
\ee
After convoluting the distribution of the spectator momentum
with the subprocess of the virtual $b$--quark decay
in a frame where its momentum is $-{\bf p}$,
we can obtain the lepton energy spectrum at the hadronic level.
This model has been extensively used in the analysis of the charged
lepton energy spectrum in semileptonic decays 
and numerically reproduce the inclusive spectra well.
In the light of QCD, several authors analyzed the model \cite{accmmh}
and suggested a way to reproduce the pattern of the QCD description of 
the heavy quark motion. 
In this paper we derive the distribution function $f(x)$ in the IMF
from the given Gaussian distribution in the $B$ rest frame, $\Phi(p)$, 
in order to study the ACCMM model 
by comparing  with the parton model approach. 

Let's choose a coordinate system where the $B$ meson is at the origin
in $B$ rest frame.
We boost the $B$ meson along the positive 
$z$ direction with the velocity ${\beta}$. 
As is well known, 
for the $B$ meson at rest we have the boosted energy of $B$ meson as
\be
E_B^{*}={\gamma} E_B= {\gamma} m_{_B} ,
\ee
where ${\gamma}=1/\sqrt{1-{\beta}^2}$, and that of $b$-quark as
\be
E_b^{*}(p)={\gamma}(E_b (p)-{\beta} p_z)  ,
\ee
where $p$ is the magnitude of the spatial momentum of the $b$--quark 
and $p_z$ is the $z$ component of ${\bf p}$ in the $B$-rest frame.
Now we calculate the energy ratio 
between $b$-quark and $B$ meson in this moving frame
\be
{\frac{E_b^{*}} {E_B^{*}}} = 
\frac{{\gamma}(E_b-{\beta} p_z)} {{\gamma} m_{_B}} .
\ee
In the limit of ${\beta} \to 1$, $\gamma \to \infty$,
we define the ratio as
\be
{\frac{E_b^{*}} {E_B^{*}}} =
\frac{(E_b-p_z)} {m_{_B}} \equiv x ,
\ee
which represents the ratio of $b$-quark  energy to
$B$ meson energy in the IMF.

Because no specific direction for ${\bf p}$ is preferred in $B$ meson decay,
the value of $p_z$ is distributed from $-p$ to $+p$
with equal probability for a fixed value of $p$.
Then the value of $x$ can be from $x_{min} (p)$ to $x_{max} (p)$ with
equal weight for a fixed $p$, where
\be
x_{min} (p) = {\frac{1}{m_{_B}}} (E_b (p) - p) ,
\nonumber \\
x_{max} (p) = {\frac{1}{m_{_B}}} (E_b (p) + p) ,
\ee
with the weight  $m_{_B}/2p$,
because  $x_{max} (p) - x_{min} (p) = 2p/m_{_B}$.
Each value of $p$ in the $B$ meson rest frame corresponds to a distribution
of $x$ in the IMF such that
the probability of finding a $b$-quark 
in a $B$ meson carrying a fraction $x$ of the meson momentum in the IMF 
is obtained by intergrating
over $p$ with Gaussian weight, $\Phi(p)$,
\be
f(x) &=& \int_{0}^{p_{max}} dp~p^2~\Phi(p)~\cdot 
\frac{m_{_B}}{2p}~\theta (x-x_{min}(p))~ \theta (x_{max}(p)-x) 
\nonumber \\
&=& \int_{p_{min}}^{p_{max}} dp~p^2~\Phi(p)~\cdot
\frac{m_{_B}}{2p}
\ee
where 
\be
p_{min} = \frac{|m_B^2 (1-x)^2 - m_{sp}^2|}
               {2 m_{_B} (1-x)}
\ee
is obtained by the conditions $x_{min}(p) < x < x_{max}(p)$.
We take the value of $p_{max}$ to be infinity
as far as the integration is concerned, 
because the area of phase space for 
$p > p_{max}$ is negligible.
After the integration over $p$, we obtain
\be
f(x) = \frac{m_{_B}}{\sqrt{\pi} p_{_F}}
       \exp \left( -\frac{1}{4} 
        \left( \frac{\rho p_{_F}}{m_{_B}(1-x)}
           -\frac{m_{_B}}{p_{_F}}(1-x)
        \right)^2
       \right)~~,
\ee
where $\rho = m_{sp}^2/p_{_F}^2$.
Using the Eq. (12) and the relation, 
$(1-x_{+}) = \frac{\bar{\Lambda}}{m_{_B}} (1-x_{_B})$, 
we find Eq. (30) becomes Eq. (30) of Ref. \cite{bigi}, 
the light--cone distribution function for the ACCMM model 
derived by Bigi {\it et al}.
We show the functional behavior of $f(x)$  in Fig. 3
by varying the parameters $p_{_F}$ and $m_{sp}$.
Note that the shape of the function $f(x)$ depends on 
two parameters $m_{sp}$ and $p_{_F}$.
The peak position is determined by $m_{sp}$:
\be
x_{peak}=\frac{m_{_B}-m_{sp}}{m_{_B}} ,
\ee
while the Fermi motion parameter $p_{_F}$ is related to the width of
the distribution $f(x)$, i.e. the larger the value of $p_{_F}$,
the broader the distribution $f(x)$.
One can easily check the normalization of $f(x)$
\be
\int_{0}^{1} f(x) dx  
 = \int_{0}^{p_{max}} dp~ p^2 \Phi(p) 
 = 1 .
\ee
The first moment of $f(x)$ is directly related to 
the average $b$-quark energy in the $B$ rest frame,
\be
\langle x \rangle \equiv \int_{0}^{1} dx~x f(x) =
\int_{0}^{p_{max}} dp~p^2 \Phi(p) \frac{E_b(p)}{m_{_B}} 
\equiv \frac{\langle E_b(p) \rangle_{B-rest}}{m_{_B}} ;
\ee
because $m_b \equiv \langle x \rangle m_{_B}$ in the parton model approach
we get the relation
$m_b \equiv \langle E_b \rangle_{B-rest}$, which 
reminds us of Cs$\acute{\mbox{a}}$ki and Randall's 
definition of the $b$--quark mass\footnote{
Csaki and Randall \cite{accmmh} showed that
the predictions of the ACCMM model agree well with those of the
heavy quark effective theory with a definition of the $b$-quark mass, 
$m_b \equiv \langle E_b(p) \rangle_{B-rest}$.}
for the ACCMM model with respect to the HQET.

We thus have derived a distribution function $f(x)$ of Eq. (28) for the parton model
approach from a Gaussian distribution in the $B$ rest frame. 
Following the method discussed in section 2,
we can calculate the moments of the line shape function
and the parameters $\L$ and  $\mu_\pi^2$
for given model parameters.
First we calculate $\L$ :
\be
\L \equiv m_{_B} - m_b 
  &=& m_{_B} - \langle E_b \rangle_{B-rest} \\ \nonumber
  &=& m_{_B} - 
     \int_{0}^{\infty} dp ~p^2 \Phi(p) (m_{_B}-\sqrt{p^2+m_{sp}^2}) \\ \nonumber
  &=& \langle \sqrt{p^2+m_{sp}^2} \rangle_{B-rest} .
\ee
{}From Eq. (14), $\mu_{\pi}^2$ is given as
\be
\mu_\pi^2 = 3 \L^2 a_2 
          = 3 m_{_B}^2 \left( \langle x^2 \rangle 
                     - \frac{m_b^2}{m_{_B}^2} \right) .
\ee
And using
\be
\langle x^2 \rangle &=& \int_{0}^{1} dx~ x^2 f(x) 
\\ \nonumber
&=& \frac{1}{m_{_B}^2}
(\langle E_b^2 \rangle_{B-rest} + \frac{1}{3} \langle p^2 \rangle_{B-rest}) ,
\ee
we get
\be
\mu_\pi^2 = 3(\langle E_b^2 \rangle_{B-rest}
- \langle E_b \rangle^2_{B-rest})
+ \langle p^2 \rangle_{B-rest} .
\ee
Finally with the following relations,
\be
\langle E_b^2 \rangle_{B-rest}
-\langle E_b \rangle_{B-rest}^2 =
\langle p^2 \rangle_{B-rest} + m_{sp}^2
- \langle \sqrt{p^2+m_{sp}^2} \rangle_{B-rest}^2
\ee
and 
\be
\langle p^2 \rangle_{B-rest}
= \int_{0}^{\infty} dp~p^2 \Phi(p)~p^2 
= \frac{3}{2} p_{_F}^2 ,
\ee
we obtain
\be
\mu_{\pi}^2=6 p_{_F}^2 + 3 m_{sp}^2
-3 \langle \sqrt{p^2+m_{sp}^2} \rangle_{B-rest}^2 .
\ee
We show the parameter space of $(p_{_F}, m_{sp})$
for given $\L$ and $\mu_\pi^2$ in Fig. 4. 
Values of $(p_{_F},m_{sp})$ in the shaded region A
satisfy the condition of Eq. (17) given in the Ref. \cite{ball},
while values in the region B satisfy the condition of Eq. (17) 
given in the Ref. \cite{neubert1}.
We can determine the model parameters $p_{_F}$ and $m_{sp}$,
once we know the correct value of the HQET parameter $\mu_\pi^2$.

\section{Discussion and Conclusions}

We have studied two phenomenological models 
(QCD parton model and ACCMM model) 
for the analysis of
inclusive semileptonic $B$ decays in the context of 
the parton model approach.
In this framework, the core of modeling is the choice of 
a distribution function of the $b$--quark inside a $B$ meson.
The light--cone dominance enables us to describe
the triple differential decay rate in terms of a single function,
the light--cone distribution function up to kinematic factors,
as the form given in Eqs. (9) and (11).
The distribution function in ACCMM model comes from 
the Gaussian distribution of the 
$b$--quark momentum at the $B$ rest frame.
Bigi {\it et al.} \cite{bigi} obtained 
the light--cone distribution function
for the ACCMM model by reading off it from the photon energy spectrum
of $B \to X_{s} \gamma$ decays.
We calculate
it here by counting the probability of finding a $b$--quark 
with the boosted energy ratio $x$ between $b$--quark and $B$ meson
into the infinite momentum frame 
within the framework of the parton model.
We find that our result of Eq. (30) exactly reproduce
the light-cone distribution function given by Bigi {\it et al.},
Eq. (30) of Ref. \cite{bigi}.
This agreement is originated from the fact that
the ACCMM model treats the $b$--quark as a quasi--free particle 
at the decay rate level by classical treatment 
as we do not consider the perturbative QCD corrections in this paper.

Even though we cannot calculate the distribution function from QCD
completely, some model independent information
can be obtained with the help the HQET. 
When we expand this function with infinite number of moments,
we cannot know all the moments while the HQET give some informations
for a few moments. 
We note that since this is an singular expansion, we cannot construct 
a well--behaved function with only a few moments.
In spite of that, the information is very useful
for giving some constraints on phenomenological models.
Investigating the distribution function,
two moments of the function, corresponding to the mean
value and the width, can be related to two QCD parameters,
$\bar{\Lambda}$ (or $m_b$) and $\mu_\pi^2$.
The mean value of the function strongly depends on
$\bar{\Lambda}$ and very weakly on the $\mu_\pi^2$,
while the width on the $\mu_\pi^2$.
Actually the fact that $a_1=0$ define the quark mass,
equivalent to $\bar{\Lambda}$ and the next moment $a_2$ is directly
related to the $\mu_\pi^2$.
With these constraints, we obtain the possible regions of
model parameters for each model, as shown in Figs. 2 and 4.

Recently, it has been an important subject to obtain an accurate value
of the kinetic energy, 
$\mu_\pi^2~(\equiv -\lambda_1)$,
of the heavy quark inside $B$ meson in connection with the heavy quark 
effective theory (HQET).
Neubert derived \cite{neubert2}
the field-theory version of the virial theorem within the HQET framework
and obtained the result $\mu_\pi^2 \sim 0.10 \pm 0.05\ {\rm GeV}^2$, 
which is much smaller than the QCD sum rule calculations 
of Ball and Braun \cite{ball}
and rather comparable to the earlier QCD sum rule result
using a less sophisticated approach in Ref. \cite{eletsky}.
However, it should be noted that Refs. \cite{ball} and \cite{neubert1}
differ in the choice of the 3-point correlation functions used to
estimate the matrix elements of interest.
The difference in the numerical values obtained in these two
calculations is understood in terms of the contributions of
excited states, which in principle must be subtracted in any
QCD sum rule analysis.
In practice, this subtraction can only be done approximately.
Therefore, the numerical differences between Refs. \cite{ball} and
\cite{neubert1} indicate the limited accuracy of the QCD sum rule
approach.
Bigi $et$ $al.$ \cite{bigi1} derived an inequality between the
expectation value of the kinetic energy of the heavy quark
inside the hadron and that of the chromomagnetic operator,
$\langle {\bf p}^2 \rangle \,\ge\, {3 \over 4} ({M_V}^2-{M_P}^2)$,
which gives $\mu_\pi^2 \ge 0.36\ {\rm GeV}^2$ for $B$ meson system.
However,
Kapustin $et$ $al.$ \cite{kapustin} showed later that this lower bound
could be significantly weakened by higher order perturbative corrections.
By means of a QCD relativistic potential model, the value of $\mu_\pi^2$
was found to be quite large as $0.44 \sim 0.46$ GeV$^2$ in Ref. \cite{hkn}.
Besides the above theoretical calculations of $\mu_\pi^2$,
Gremm $et$ $al.$ \cite{gremm} extracted the average kinetic energy
by comparing the prediction of the HQET with
the shape of the inclusive $B \rightarrow X l {\nu}$
lepton energy spectrum \cite{cleo93b} for $E_l \ge 1.5$ GeV,
in order to avoid the contamination from the secondary leptons of cascade
decays of $b \rightarrow c \rightarrow s l \nu$.
They obtained
$-\lambda_1 = 0.19 \pm 0.10\ {\rm GeV}^2$.
Combining the experimental data on the inclusive decays of 
$D \rightarrow X e \nu$, $B \rightarrow X e \nu$ and
$B \rightarrow X \tau \nu$,
Ligeti $et$ $al.$ \cite{Ligetiexp} derived the bound of 
$\mu_\pi^2 \le 0.63$ GeV$^2$ if $\bar \Lambda \ge 0.240$ GeV, or
$\mu_\pi^2 \le 0.10$ GeV$^2$ if $\bar \Lambda \ge 0.500$ MeV.
Li $et$ $al.$ \cite{Li} obtained the value of $-\lambda_1$
centered at $0.71$ GeV$^2$ from the
analysis of the inclusive radiative decay
$B \rightarrow X_s \gamma$ \cite{Liexp} within the perturbative
QCD framework.
Related with the comparison of various theoretical calculations of
$\mu_\pi^2$, we note that Ref. \cite{neubert3} emphasizes that one has to
be careful when comparing the values of $-\lambda_1$ obtained using
different theoretical methods.
The mixing of the operator for the heavy quark kinetic energy
with the identity operator occurs at the two--loop order
and the parameter $\lambda_1$ requires a non-perturbative subtraction.

Note also that the functional behavior of Refs. \cite{kylee}
and \cite{accmm} is different from each other
while both functions satisfy the constraints. 
Even though two moments, $a_1$, $a_2$, which correspond
to the mean values and the widths of the distribution functions,
agree with each other, their full behaviors do not necessarily agree.
For the two models studied in this paper,
we find that the modified Peterson's function in Fig. 1
has a rather longer tail than the distribution
function derived from the Gaussian function of the Fermi momentum
defined in the ACCMM model in Fig. 3.
We cannot say which model is better, at least at this stage.
Both models can give proper values for the $b$--quark mass
and the width of the charged lepton energy spectrum,
both theoretically and phenomenologically.
For further tests, we must fit them to the experimental data
more precisely after more data is accumulated.
For this purpose the photon energy spectrum of 
the inclusive radiative decay $B \to X_s \gamma$ 
would be more useful. 
The photon energy spectrum of this process is known to be 
more sensitive to the behavior of the distribution function 
than the charged lepton energy spectrum of 
the inclusive semileptonic decays \cite{kylee,Li},
since the charged lepton energy spectrum is contaminated by kinematics.

\acknowledgements

We would like to thank M. Drees for useful suggestions and helpful comments.
The work of CSK was supported 
in part by Non-Directed-Research-Fund, KRF in 1997,
in part by the CTP, Seoul National University, 
in part by Yonsei University Faculty Research Fund of 1997, 
in part by the BSRI Program, Ministry of Education (MOE), 
Project No. BSRI-97-2425,
and in part by the KOSEF-DFG large collaboration project, 
Project No. 96-0702-01-01-2.
YGK was supported by CTP of SNU.
KYL is a postdoctoral fellow supported by Korea Research Foundation
(KRF) and Research University Fund of College of Science
at Yonsei University by MOE of Korea.

%
%
\newpage
\begin{figure}
\caption{
(a) Functional behavior of the improved Peterson's function
with varying $\alpha_{_Q}$ = 0.05, 0.085, 0.1 for fixed $\epsilon_{_Q} = 0.004$;
(b) the same function with varying $\epsilon_{_Q}$ = 0.002, 0.004, 0.006 
for fixed $\alpha_{_Q} = 0.085$.
}
\end{figure}
\begin{figure}
\caption{
The model parameter space in the ($\alpha_{_Q}$, $\epsilon_{_Q}$) plane
for the QCD sum rule results for $\L$ and $\mu^2_{\pi}$.
Values of ($\alpha_{_Q}, \epsilon_{_Q}$) in the shaded region
satisfy the condition $0.4 < \L < 0.6$ and
$0.5 < \mu_\pi^2 < 0.7$.
}
\end{figure}
\begin{figure}
\caption{
(a) Functional behavior of the distribution function
in the infinite momentum frame derived from the Gaussian
distribution of spectator momenta
with varying $p_{_F}$ = 0.3, 0.5, 0.7 GeV for fixed $m_{sp}$ = 0.1 GeV;
(b) the same function with varying $m_{sp}$ = 0.0, 0.2, 0.4 GeV 
for fixed $p_{_F}$ = 0.5 GeV 
}
\end{figure}
\begin{figure}
\caption{
The model parameter space in the ($p_{_F}$, $m_{sp}$) plane
for the QCD sum rule results for  $\L$ and $\mu^2_{\pi}$.
Values of ($p_{_F}, m_{sp}$) in the shaded region A
satisfy the condition $0.4 < \L < 0.6$ and
$0.5 < \mu_\pi^2 < 0.7$.
Values of ($p_{_F}, m_{sp}$) in the shaded region B
satisfy the condition $0.4 < \L < 0.6$ and
$0.05 < \mu_\pi^2 < 0.15$.
}
\end{figure}

\newpage

\begin{figure}[th]
\centering
\centerline{\epsfig{file=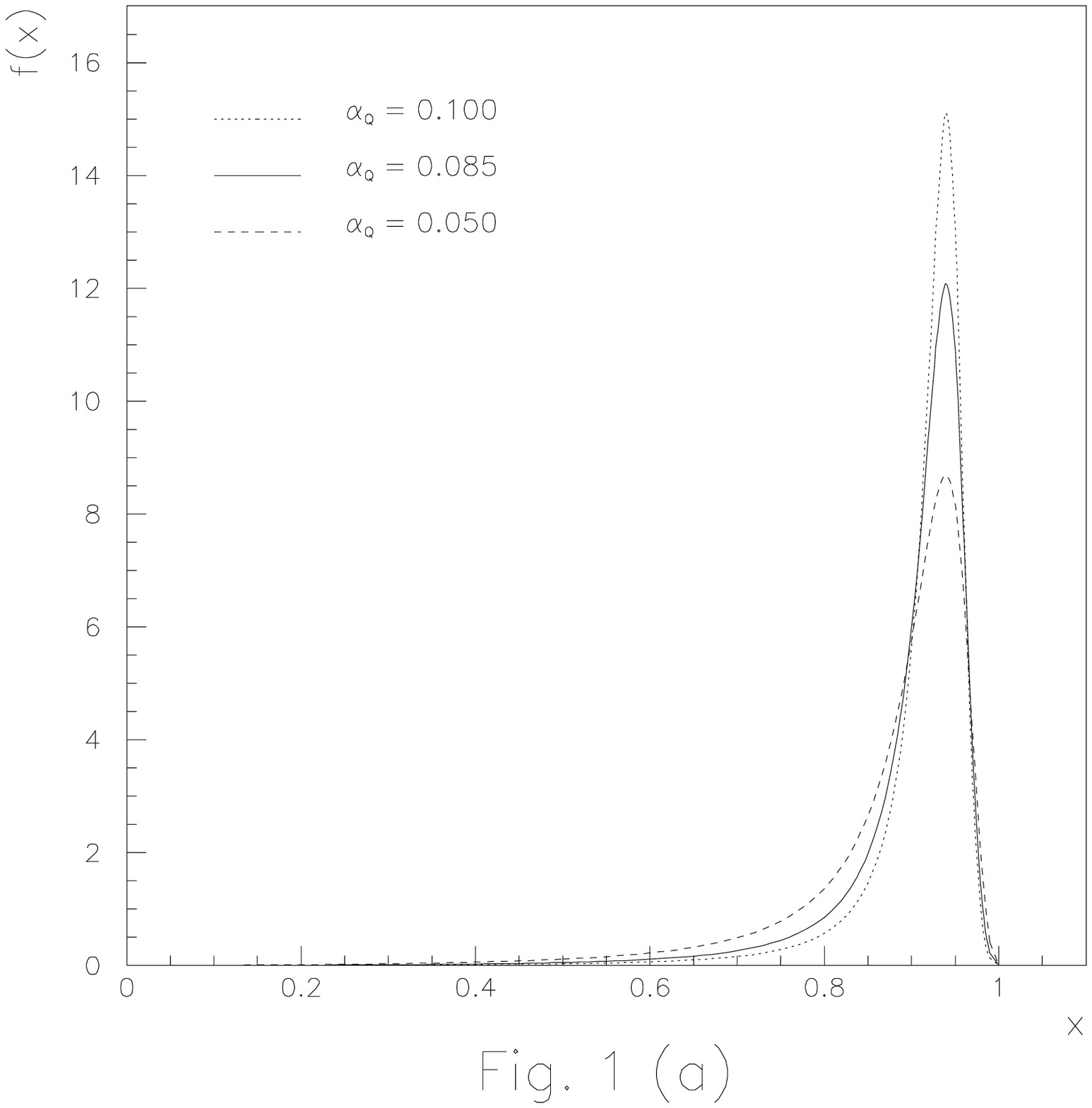}}
\end{figure}

\begin{figure}[th]
\centering
\centerline{\epsfig{file=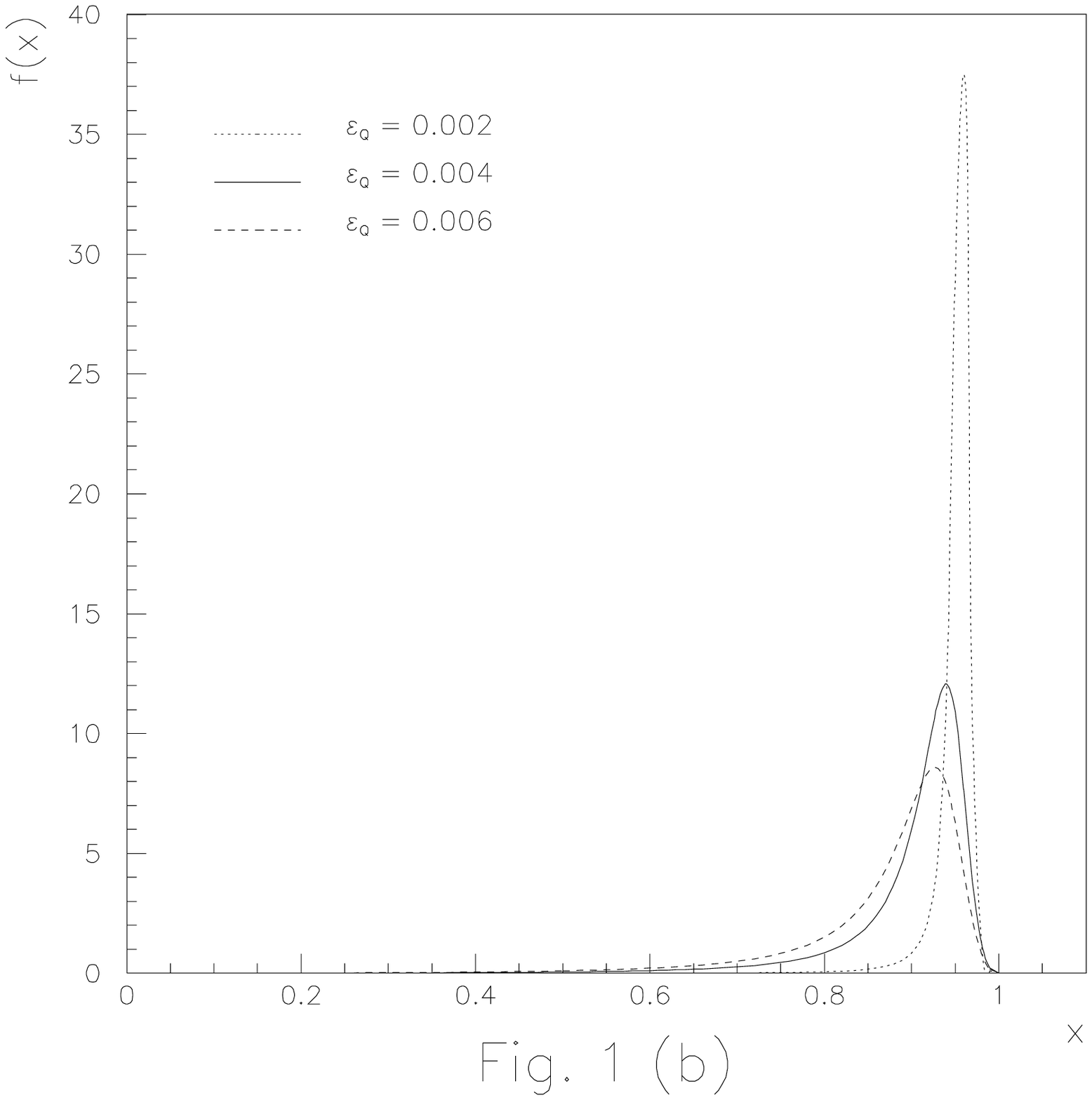}}
\end{figure}

\begin{figure}[th]
\centering
\centerline{\epsfig{file=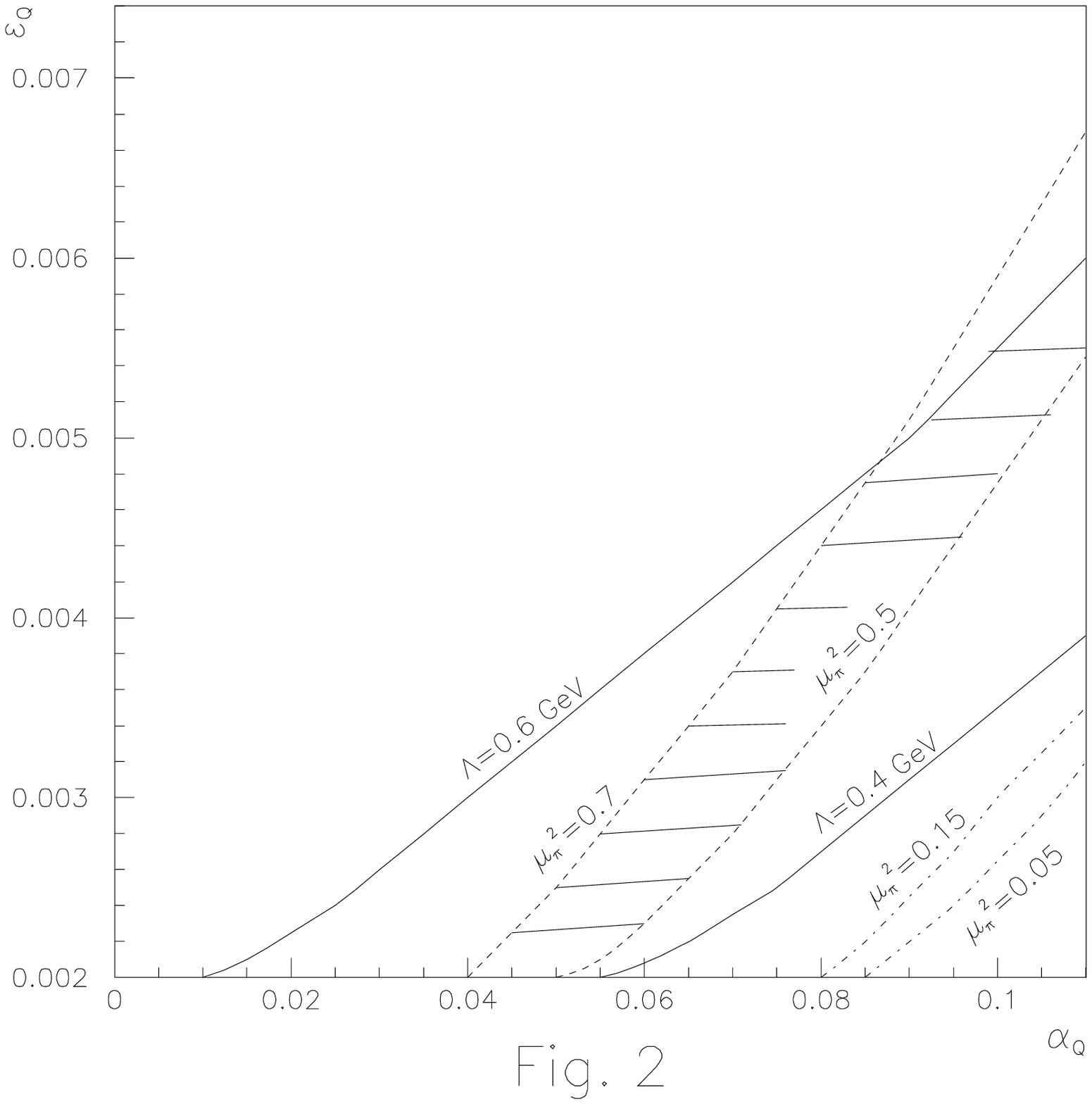}}
\end{figure}

\begin{figure}[th]
\centering
\centerline{\epsfig{file=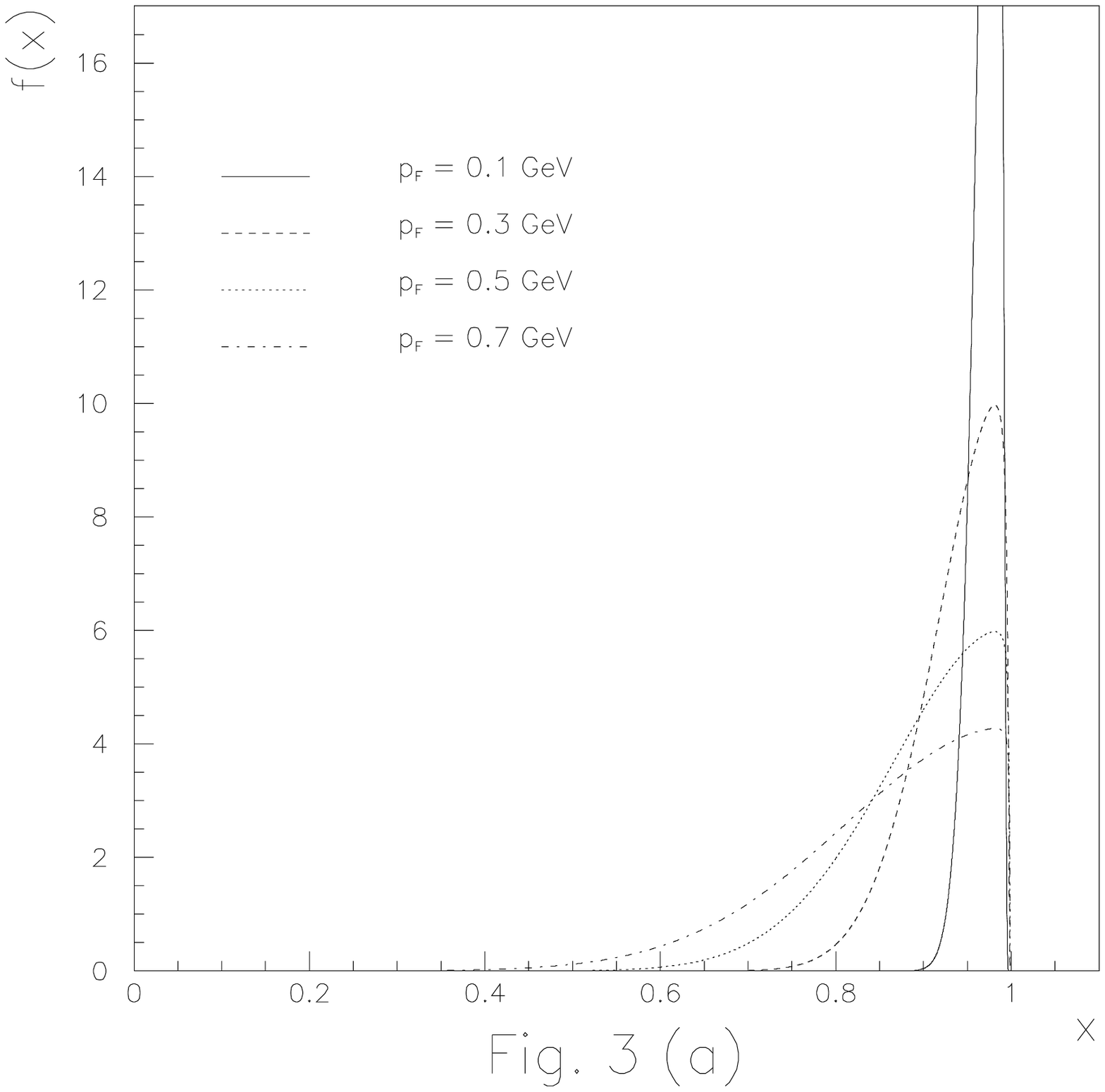}}
\end{figure}

\begin{figure}[th]
\centering
\centerline{\epsfig{file=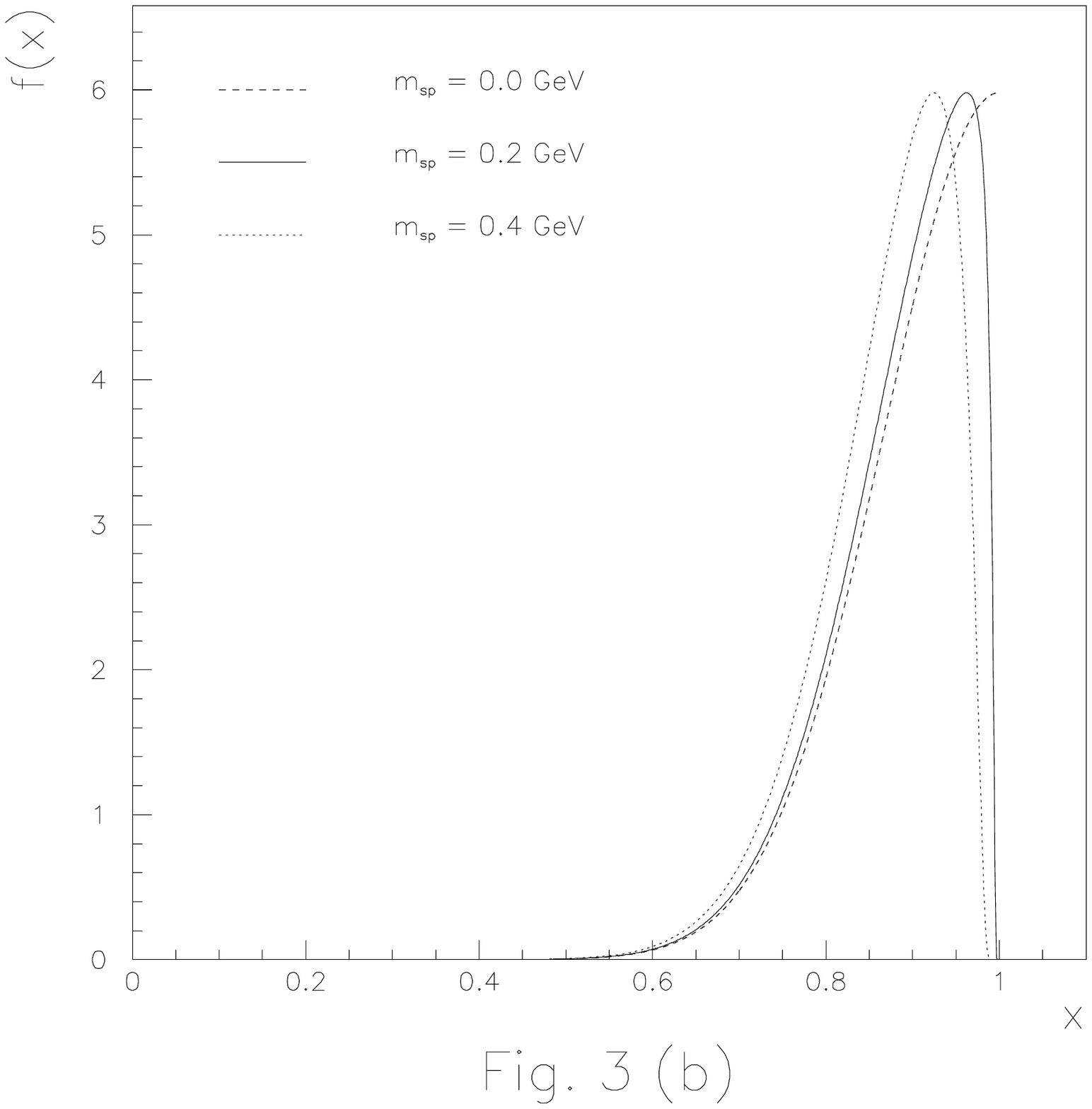}}
\end{figure}

\begin{figure}[th]
\centering
\centerline{\epsfig{file=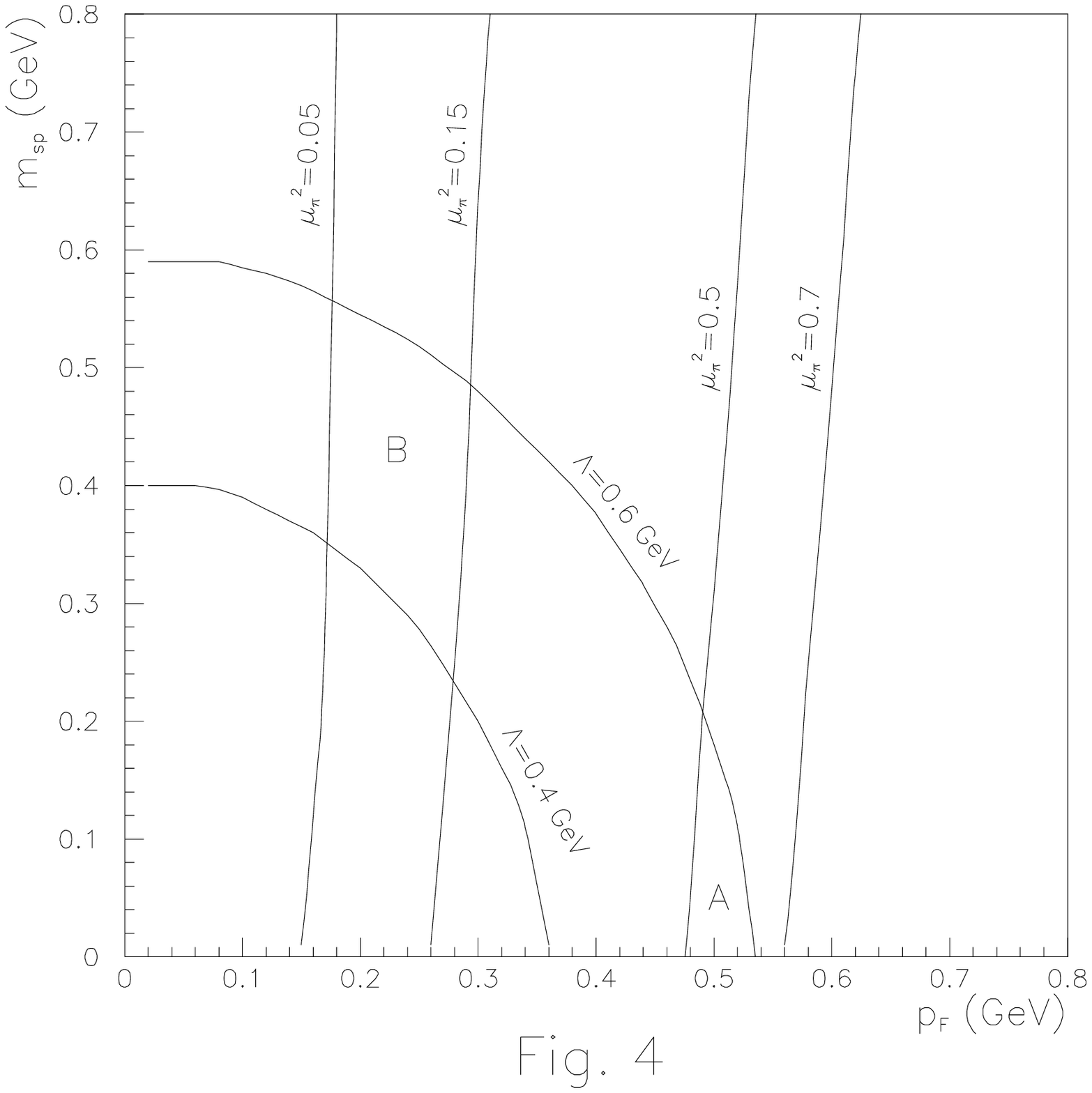}}
\end{figure}

\end{document}